\documentclass[final,5p,times,twocolumn]{elsarticle}
\usepackage{graphicx}
\usepackage{amssymb}
\usepackage{amsmath}
\usepackage{aas_macros}
\usepackage{hyperref}
\journal{Astroparticle Physics}
\begin{document}
\begin{frontmatter}

\title{The Energy Spectrum of Ultra-High-Energy Cosmic Rays Measured
by the Telescope Array FADC Fluorescence Detectors in Monocular Mode}

   \author[label1]{T.~Abu-Zayyad}
   \author[label2]{R.~Aida}
   \author[label1]{M.~Allen}
   \author[label1]{R.~Anderson}
   \author[label3]{R.~Azuma}
   \author[label1]{E.~Barcikowski}
   \author[label1]{J.~W.~Belz}
   \author[label1]{D.~R.~Bergman}
   \author[label1]{S.~A.~Blake}
   \author[label1]{R.~Cady}
% ...

   \address[label1]{University of Utah, High Energy Astrophysics Institute, Salt Lake City, Utah, USA}
   \address[label2]{University of Yamanashi, Interdisciplinary Graduate School of Medicine and Engineering, Kofu, Yamanashi, Japan}
   \address[label3]{Tokyo Institute of Technology, Meguro, Tokyo, Japan}
% ...

  \cortext[cor1]{Corresponding author}
  \fntext[email]{tstroman@physics.utah.edu}

\begin{abstract}
We present a measurement of the energy spectrum of ultra-high-energy
cosmic rays performed by the Telescope Array experiment using
monocular observations from its two new FADC-based fluorescence detectors.
After a short description of the experiment, we describe
the data analysis and event reconstruction procedures. Since the
aperture of the experiment must be calculated by Monte Carlo
simulation, we describe this calculation and the comparisons of
simulated and real data used to verify the validity of the aperture
calculation. Finally, we present the energy spectrum calculated
from the merged monocular data sets of the two FADC-based detectors,
and also the combination of this merged spectrum with 
an independent, previously published monocular 
spectrum measurement performed by Telescope Array's 
third fluorescence detector %\cite{2012APh....39..109A}.
(Abu-Zayyad {\it et al.}, {Astropart. Phys.} 39 (2012), 109).
This combined spectrum corroborates the recently published
Telescope Array surface detector spectrum %\cite{2013ApJ...768L...1A} 
(Abu-Zayyad {\it et al.}, {Astrophys. Journ.} 768 (2013), L1)
with independent systematic uncertainties.
\end{abstract}


\end{frontmatter}
\begin{figure}[tb]
\includegraphics[width=3in]{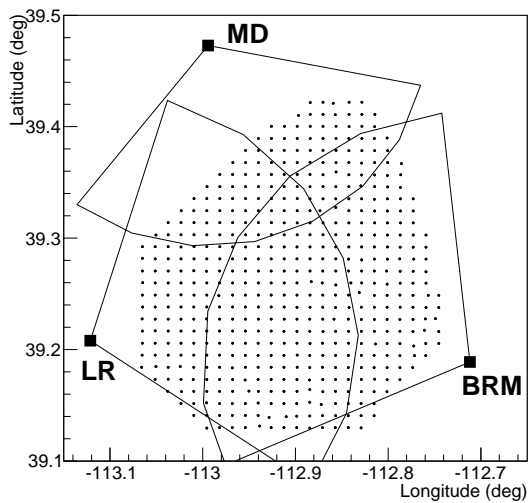} %ta-layout-with-labels.eps
\caption{\label{fig:ta-layout}
The Telescope Array experiment 
in western Utah consists of three FDs 
(squares) and 507 SDs (dots). The southeastern and southwestern FDs 
(Black Rock Mesa and Long Ridge, respectively) 
use new telescopes designed for Telescope
Array and based on FADC electronics; the northern FD (Middle Drum) consists of
telescopes refurbished from the High Resolution Fly's Eye experiment.
The FDs' approximate fields of view for 
$10^{19}$-eV cosmic rays are outlined. The average distance between
neighboring SDs is 1200~m.
}
\end{figure}

\section{Introduction}
Ultra-high-energy cosmic rays (UHECRs) are charged subatomic particles 
of extraterrestrial origin with kinetic energies above
$10^{18}$~eV, making them the most energetic 
particles in the known universe. 
A clear understanding of their origins and 
chemical composition has not yet been experimentally established, 
largely due to the scarcity of UHECRs: 
collecting enough data to suppress uncertainty from 
small-number statistics requires a detector that can observe 
a large area for a long time. 
The Telescope Array (TA) experiment in western Utah is the 
largest UHECR detector currently operating in the northern 
hemisphere\footnote{The largest UHECR detector is the Pierre Auger Observatory in Argentina \cite{2004NIMPA.523...50A}.}
\cite{2008ICRC....4..417F,Matthews-2009-ICRC-31-1386}. 
Centered at approximately 112.9$^\circ$~W, 39.3$^\circ$~N 
near the city of Delta in Millard County, 
TA is a ``hybrid'' detector consisting of three atmospheric 
fluorescence detector (FD) stations and a ground array of 507 surface 
detectors (SDs) on a square grid with 1200-meter spacing 
(see Figure~\ref{fig:ta-layout}).

Each of the FDs and the SD array operate independently, 
collecting data for UHECR measurements. 
The SDs, which directly detect secondary particles in the 
extensive air shower produced by a primary UHECR, 
collect data night and day in all weather and thus 
have a duty cycle of nearly 100\%. 
The FDs use telescopes to measure ultraviolet light produced 
when an air shower excites atmospheric N$_2$. 
For best sensitivity, 
FDs operate only on moonless nights,
so their duty cycles are each approximately 10\%. 

Although the SD array alone boasts the strongest statistical power within
the experiment, 
combining and comparing data from different components of TA allows 
distinct, corroborating measurements of physical quantities of interest.
Using the simultaneous observation of a single cosmic-ray air shower by one FD and
either the SD array (``hybrid'') or a second FD (``stereo''), 
we tightly constrain certain
geometric properties of the air shower, but the majority of UHECRs do not satisfy
this observation criterion. The ``monocular'' observation of UHECRs,
reconstructing events using measurements from a single FD
station, accumulates data at a rate second only to the SD array, and has several
additional advantages over hybrid or stereo analysis: it encompasses a broader
range of UHECR energies, its aperture calculation is less sensitive to 
atmospheric variation than the corresponding stereo calculation, 
and its systematic uncertainties are independent of those
used in the SD analysis. A monocular measurement of the energy spectrum
is therefore an important complement to the same spectrum as measured by
the SD array.

Two different designs of FD station are in use at TA: 
the northern station, Middle Drum (MD), uses refurbished hardware from the
High Resolution Fly's Eye (``HiRes'') cosmic-ray experiment, 
which collected data at Utah's Dugway Proving Ground from 
1997 to 2006 \cite{2008PhRvL.100j1101A}. 
MD's data acquisition (DAQ) system is based on
sample-and-hold electronics, 
in which each pixel of a telescope's image reports a single
value for signal intensity and a time reference. 
The southeastern and southwestern FD stations, 
respectively dubbed Black Rock Mesa (BRM) and Long Ridge (LR), 
consist of new telescopes designed for TA that use 
flash analog-to-digital converter (FADC)-based
electronics to record the evolution of each pixel's signal intensity. 

In this paper, 
we report the UHECR energy spectrum above $10^{18}$~eV
as measured by the two 
FADC-based FDs operating in monocular mode. 
The corresponding measurement by the MD FD has
been reported elsewhere \cite{2012APh....39..109A}, 
as has the energy spectrum measured 
by the SD array \cite{2013ApJ...768L...1A}. 
In Section~\ref{sec:fadc}, 
we elaborate on the construction and operation of the BRM and LR FDs, 
whose data we analyze as described in Section~\ref{sec:analysis}. 
Section~\ref{sec:aperture} describes the Monte Carlo simulation 
process by which we calculate the detectors' sensitivity. 
We present the energy-spectrum measurement in Section~\ref{sec:spectrum},
followed by a combination of our measurement with that 
from the MD FD (Section~\ref{sec:combine}).
We conclude with a discussion of our results in Section~\ref{sec:discussion}.
Our results are corroborated by a separate monocular analysis 
not described in detail here, 
using computer programs and processing techniques developed independently
from ours \cite{2012.fujii.isvhecri,2012.fujii.thesis}.

\section{\label{sec:fadc}FADC-based fluorescence detectors}
\begin{figure}[tb]
\includegraphics[width=3.4in]{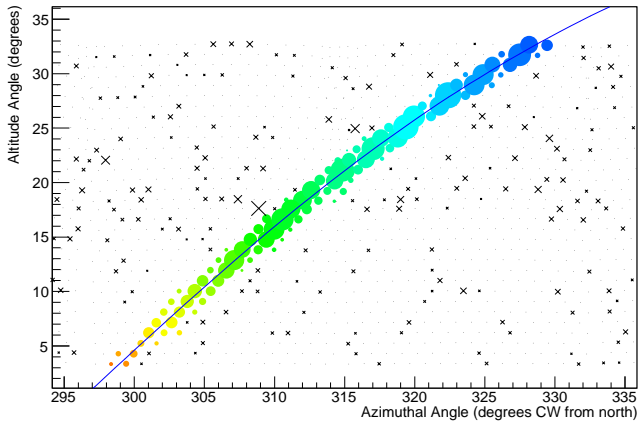} %sample-event-sky.eps
\caption{\label{fig:skymap}
The projection on the sky of the signal produced by one event in our data set.
The colored circles represent ``good'' PMTs (those that are included in the
geometry fit and subsequent Gaisser-Hillas reconstruction), where the diameter of
each circle is proportional to $N_{pe}$ and the color represents the weighted average
signal time: the earliest signals are blue, with the last signals 
(some 13~$\mu$s later in this example) colored orange. 
PMTs that are designated ``noise'' and excluded from the reconstruction are marked with
the symbol $\times$, and the viewing directions of PMTs discarded during 
raw-data preprocessing are marked with single dots. The calculated shower-detector plane
(Equation~\ref{eqn:sdp}) is superimposed as a solid line.}
\end{figure}

The TA experiment's FADC-based FDs occupy two sites at the southern
end of the array. 
The BRM FD site contains twelve telescopes with a contiguous field of view
ranging from 3$^\circ$ to 33$^\circ$ in elevation in directions to the
west and northwest (as shown in Figure~\ref{fig:ta-layout}).
The LR FD site is identical to BRM in its construction, 
but with an eastward orientation.
Each telescope consists of a segmented spherical
mirror 3.3 meters in diameter, 
which focuses light from a 15-degree (elevation)
by 18-degree (azimuth) region of the sky onto a 
cluster of 256 hexagonal photomultiplier
tubes (PMTs; Hamamatsu model R9508) \cite{2012NIMPA.676...54T}. 
The PMTs are sampled by FADC electronics at an
effective rate of 10~MHz with a 14-bit dynamic range, 
tracking the mean and variance ($\sigma^2$) over sliding time 
windows of 0.8, 1.6, 3.2, 6.4, and 12.8 $\mu$s. 
When a given PMT's
instantaneous signal minus the mean in any time window 
exceeds six standard deviations ($6\sigma$), 
the trigger criterion for that PMT is met. 
Five or more contiguous triggered
PMTs within a 25.6-$\mu$s trigger frame result 
in a readout of a 51.2-$\mu$s waveform from
all 3072 PMTs (the trigger frame plus a 12.8-$\mu$s buffer 
immediately before and after) \cite{2009NIMPA.609..227T}. 
A 12.8-$\mu$s overlap between consecutive trigger frames ensures
continuous detector acceptance during the frame transition. The typical
trigger rate for a single FD site is approximately 2~Hz.

A typical night of FD operation is broken into several parts, 
in which each part consists of
a predetermined number of triggers. 
Data parts may last from a few minutes to more than an hour. 
Parts for data collection are
interspersed with parts designated for calibration purposes, 
which may involve reduced trigger thresholds 
or the use of an artificial light source that renders those parts
unsuitable for data analysis. 
For much of the observation time reported here, 
the LR FD site has been operated remotely from the
control room at the BRM FD site; 
the logistics of remote operation have resulted in a smaller duty
cycle for LR than for BRM.

We assign a single numerical score to describe the weather 
on each night of operation based on
human FD operators' logs as well as automated monitoring, and
designate adjacent ranges of scores as ``excellent,'' ``good,'' 
and ``bad'' weather. In this analysis, 
we use only data from nights with good or excellent weather scores.
In addition to rejecting data from nights with
bad weather scores, we reject data parts that have 
a low rate of downward-going events whose average angular speed 
is less than $40~^\circ~\mu{\rm s}^{-1}$.
The typical rate of such events in good weather is 0.1 Hz;
we reject parts with rates below 0.067 Hz.
Our results include
data taken from 
the beginning of April 2008 through mid-September 2011, representing 2020.8~h eligible gross night-sky time for BRM and 1835.5~h for LR.
The readout of a detector following
a trigger results in a brief period of insensitivity. 
The average ``dead time'' fraction for
BRM and LR reduces their observing times by 7.8\% and 8.7\% 
of their respective on-times. 
These on-time values do not include data parts 
excluded from the analysis for reasons related to
DAQ errors or lack of 
instantaneous detector calibration information.

Calibration of the FDs 
(described in detail elsewhere \cite{2009NIMPA.601..364T}) 
is based on the absolute calibration of a subset of PMTs, 
two or three per telescope, in the laboratory. 
A controlled quantity of 337.1-nm
photons illuminate the photocathode of a 
PMT connected to DAQ electronics identical
to those in the field, including cables of the correct length, 
to determine the precise relationship 
between electronic response and incident light \cite{2012NIMPA.681...68K}. 
To detect any long-term calibration drift, 
these PMTs are outfitted with a small amount of $\alpha$-emitting
material adjacent to a scintillator to 
provide a reference signal. 
Additionally, each telescope is equipped with a 
xenon flasher at the 
center of the mirror that illuminates that telescope's 
entire PMT cluster over the course of a night of observation, 
which enables the accurate propagation of the 
reference PMTs' calibration to the remainder of the cluster,
as well as tracking variations in PMT gain 
with changing ambient temperature. 
The wavelength-dependent reflectance of the mirrors is measured monthly, 
with values interpolated for every ten-day period. 
Two additional wavelength-dependent quantities are assumed to be constant, 
the transparencies of the ``BG3'' UV-bandpass filter affixed to each PMT and
of the Paraglas acrylic window installed as a 
protective cover over the entire PMT cluster. 
This combination of time-independent and time-dependent 
detector calibrations enables us to accurately
interpret the recorded electronic signals in terms of 
incident photoelectrons with hourly time
resolution during the data-analysis stage.

\section{\label{sec:analysis}FD data analysis}
\begin{figure}[tb]
\includegraphics[width=3.4in]{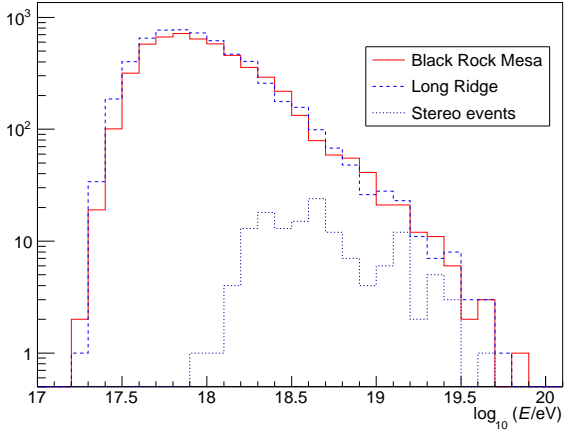} %energy_histograms.eps
\caption{\label{fig:ehist}
  The distribution of reconstructed primary
energies (without cut on energy) of cosmic rays detected by Black Rock Mesa (solid red line)
and Long Ridge (dashed blue line). 
Entries appearing in both distributions simultaneously are also
shown (dotted line) by the geometric mean of their two 
monocular reconstructed energies.}
\end{figure}

Analyzing the data collected by the FDs is a process of several steps,
beginning with the raw data and ending with a set of 
cosmic-ray events whose trajectories and air-shower longitudinal 
profiles satisfy quality cuts carefully chosen via 
Monte Carlo simulation. Steps of data reduction
alternate with steps of further processing.

Our first step in data analysis is preprocessing 
to remove unwanted PMTs: those mirrors that aren't neighbor 
to a triggered mirror are discarded, and those PMTs within the
retained mirrors whose signal never attains a $3\sigma$ 
excess above the baseline are also discarded.
The remaining PMTs, which consist of true ``signal'' PMTs as 
well as ``noise'' PMTs whose background fluctuations 
reached $3\sigma$ significance, 
are eligible for consideration in all further stages of analysis, 
which begins with geometry reconstruction.

Every PMT's waveform is processed by a digital signal processing 
algorithm that subtracts the reported background level, 
calibrates and integrates the remaining signal (number of 
photoelectrons $N_{pe}$), 
and calculates the weighted average arrival time at that PMT.
The shower-detector plane (SDP) is defined by the straight-line trajectory of 
the UHECR and a single reference point at the FD site. 
Each PMT $i$ has a nominal pointing direction $\hat{v}_i$ 
relative to the FD site, 
and the SDP is determined by an iterative process to be
that whose normal vector $\hat{n}$ minimizes the plane-fit $\chi^2$ 
given by
\begin{equation}
\label{eqn:sdp}
\chi^2=\sum_{i=1}^{N_{\rm good}}
\left(\hat{n}\cdot\hat{v}_i\right)^{2}N_{pe,i},
\end{equation}
where $N_{\rm good}$ is the number of ``good'' PMTs 
(a clustering algorithm identifies 
high-significance PMTs contiguous with 
other high-significance PMTs, as expected for emission along a line source).
Figure~\ref{fig:skymap} illustrates one example of an event with the SDP
calculated from the weighted PMT directions.
With the SDP thus determined, 
the impact parameter $R_p$ and inclination angle $\psi$ 
of the shower track within the plane are related to 
the arrival time $t_i$ of the signal in each PMT by the equation
\cite{1985NIMPA.240..410B}
\begin{equation}
t_i = t_0 + \frac{R_p}{c}\tan\left[\frac{1}{2}\left(\pi-\psi-\alpha_i\right)\right],
\end{equation}
where $t_0$ is the time of closest approach to the FD 
and $\alpha_i$ is the viewing angle of the $i^{\rm th}$ PMT 
within the SDP relative to the direction to 
the shower-core impact location.
The three parameters ($R_p$, $\psi$, and $t_0$) are determined 
by fitting the viewing angles and signal times of good PMTs 
with central viewing directions not displaced more 
than $2^\circ$ from the SDP. 
An air shower's Moli\`ere radius subtends less than $2^\circ$ at
the typical distance ($> 5$~km) to the showers 
in our data set \cite{2012PhRvD..86a0001B}, 
so this limit excludes predominantly ``noise'' PMTs far from the SDP.

Many of the events that are reconstructed in this way 
do not preserve enough good PMTs, or have
an otherwise unfavorable geometry, 
for a high probability of accurate reconstruction; 
these events are discarded. 
A Monte Carlo FD simulation (described in detail in 
Section~\ref{sec:aperture}) 
reveals the parameter space in which the detectors' resolution 
(reconstruction accuracy) is optimized. 
We subject the reconstructed events to a series of
quality cuts developed to ensure this accuracy. 
The cuts for monocular FD reconstruction 
are described in Table~\ref{tab:gcuts}. 
  
\begin{table}[tb]
\centering
  \begin{tabular}{lcc}
    Geometry cut &&\\
    \hline
    Successful timing fit && \\
    Good PMT fraction      & $\ge$ & 3.5\%\\
    Number of good PMTs    &$\ge$& 6\\
    $N_{pe}$/degree              & $>$ & 25\\
    Distance (angular speed) & $>$ & 1.5 km\\
    SDP angle               & $\le$ & 80$^\circ$\\
    $R_p$                   & $\ge$ & 500 m\\
    $\psi$ fit uncertainty  & $<$ & 36$^\circ$\\
    Timing fit $\chi^2$/DOF & $<$ & 10\\
    Track length            & $>$ & 7$^\circ$\\
    \hspace{0.1in}(including Ring-2)    & $>$ & 10$^\circ$\\
    Zenith angle            & $<$ & 70$^\circ$\\
    $t_0$                   & $<$ & 25.6 $\mu$s\\
    $\psi$                  & $<$ & 110$^\circ$\\
    $\Delta t$ (duration)   & $>$ & 6 $\mu$s (for $R_p<5$ km)
  \end{tabular}
  \caption{\label{tab:gcuts}Cuts used to select good data after the 
  geometry fitting stage of the Black Rock Mesa and Long Ridge monocular analysis. The ``Ring-2''
  track-length cut applies to shower tracks that are detected in whole or in part by
  those telescopes with fields of view at higher elevation angles 
  ($18^\circ$ to $33^\circ$).
  }

\end{table}
Those events whose geometry and signal qualify them for 
further processing move on to the next stage of data analysis: 
shower profile reconstruction. 
The evolution of the number of charged particles $N_e$ 
in an air shower as a function of slant depth $X$ 
(traversed atmospheric column density, measured in g~cm$^{-2}$), or shower profile, 
can be described on average by the Gaisser-Hillas (GH) formula \cite{1977ICRC....8..353G},
\begin{equation}N_e\left(X\right)=N_{\rm max}
\left(\frac{X-X_0}{X_{\rm max}-X_0}\right)^\frac{X_{\rm max}-X_0}{\Lambda}
\exp\left(\frac{X_{\rm max}-X}{\Lambda}\right).
\end{equation}
The four GH parameters ($N_{\rm max}$, $X_{\rm max}$, $X_0$, 
and $\Lambda$) respectively represent the 
maximum number of charged particles in the shower, 
the slant depth of shower maximum, 
an offset in slant depth, 
and the characteristic interaction depth between 
subsequent particle generations within the shower. 
Because we don't typically observe the beginning or end of
the shower, we are insensitive to $X_0$ and $\Lambda$,
so we have fixed these parameters to values based on
averages of fits to showers generated by the 
cosmic-ray simulator CORSIKA \cite{1998cmcc.book.....H}: 
$X_0 \equiv -100$~g~cm$^{-2}$ and $\Lambda \equiv 60$~g~cm$^{-2}$. 
We reconstruct the GH profile of a particular shower 
by simulating the detector's response to a shower with
a given set of GH parameters, 
and varying those parameters until the best agreement between the simulated
and observed time-integrated numbers of photoelectrons on a 
tube-by-tube
level is attained.

To simulate the FD signal produced by an air shower, 
we must determine, in sequence,
the amount of energy deposited
into the atmosphere by the air shower, 
the production of light in response to that energy deposit,
the attenuation of that light en route to the detector, 
the acceptance of light by the detector optics, 
and the response of the electronics to the accepted light. 
The atmospheric temperature, pressure, and density profiles 
are obtained from nightly radiosonde measurements at the Salt Lake
City International Airport (SLC). 
We first divide the trajectory into 
longitudinal segments of equal slant depth $\Delta X=1$~g~cm$^{-2}$. 
The value of $N\left(X\right)$ at the center of
each segment is translated to an atmospheric energy deposit 
for that segment according to Nerling {\it et al.} \cite{2006APh....24..421N}.
The fluorescence yield for that energy deposit is 
distributed according to the spectrum measured by the FLASH collaboration
\cite{2008APh....29...77A} in bins of $\Delta\lambda=5$~nm, 
with overall normalization determined by Kakimoto {\it et al.} \cite{1996NIMPA.372..527K}. 
Isotropic emission of fluorescence photons
originates from points uniformly distributed 
along the length of the segment, 
with a radial displacement distribution 
given by the NKG function 
\cite{1958PThPS...6...93K,Greisen-1956-ProgCRPhys3}. 
In addition to fluorescence light,
we calculate the production of Cherenkov radiation 
along the shower trajectory.

Light emitted toward the FD from points along the air shower 
will be attenuated as it traverses the atmosphere. 
Additionally, 
the atmosphere along the shower may scatter light toward the FD.
We calculate the effects on radiation transport from three mechanisms: 
absorption by ozone,
Rayleigh scattering by the molecular 
atmosphere via the SLC radiosonde measurements, 
and Mie scattering by atmospheric aerosols. 
In our analysis, 
we use a time-averaged aerosol distribution based on
{\it in situ} measurements at the TA site \cite{2011NIMPA.654..653T}. 
The use of an average aerosol distribution has been shown to introduce no energy-dependent effects in event reconstruction given the generally low level of aerosols in the Utah west desert \cite{2007APh....27..370H}.
The aerosol density decreases exponentially
with height over a scale length of 1~km, 
normalized to reproduce the median vertical aerosol optical depth 
(VAOD) of 0.034 at 1370~m altitude.

The FD acceptance of incident light is calculated by ray tracing. 
Each emitting segment along the shower track within $15^\circ$ 
of the center of a particular telescope's field of view is simulated, 
with photons directed into a circle circumscribing 
the 3.3-m diameter mirror. 
Rays that are obstructed by the PMT cluster or its support structure, 
or which pass between mirror segments, 
or which reflect but do not land on the face of any PMT, are discarded. 
Additionally,
the measured non-uniformity of the PMTs' sensitivity implies that 
even rays incident on a PMT's photocathode may not be accepted. 
The accepted light fraction is oversampled by a factor
of 10 (20 in the final calculation) over the observed flux to 
ensure that the acceptance from faint segments 
is not dominated by small-number statistics.

Photoelectrons produced in the PMT photocathode by accepted light 
produce a signal in the DAQ system. 
The precise time of photoelectron arrival within a 100-ns waveform bin
determines the relative distribution of the signal among 
that bin and its successors via the impulse response function. 
The signals from all of the accepted photoelectrons,
as well as a measured background contribution 
(9 photoelectrons per 100~ns on average), 
are superimposed to produce a single simulated waveform in each PMT. 
The integrated signal from the simulated waveform is 
compared with the value of the observed event in the data.

The initial value of $N_{\rm max}$ for each shower's simulation is chosen 
using a crude estimate from the shower geometry and signal intensity, 
and $X_{\rm max}$ begins at 750~g~cm$^{-2}$. 
After the initial acceptance calculation, 
the best-estimate number of charged particles in each track segment is calculated 
from the observed signal using that acceptance and the 
atmospheric transparency, and a GH profile is fit to the estimate. 
If the fit value of $X_{\rm max}$ changes from the simulation value 
by more than 50~g~cm$^{-2}$, 
the acceptance is recalculated. 
This process may be repeated additional times until the fit $X_{\rm max}$ is within
50~g~cm$^{-2}$ of the value used in the most recent acceptance calculation, 
at which time a final acceptance calculation 
and fit are performed using a higher
oversampling factor than the preparatory calculations.

We calculate the calorimetric energy $E_{\rm cal}$ of the primary UHECR by 
integrating the product of the fit GH profile and the ionization 
loss rate along the entire shower track. 
The ``missing energy,'' the portion of the primary's energy 
going into muon and neutrino production, 
is added to the calorimetric energy in an amount determined 
by CORSIKA; 
the missing energy accounts for 7\%--10\% of the total UHECR energy, 
$E_0$ \cite{2012.stratton.thesis}:
\begin{eqnarray}
\nonumber \frac{E_{\rm cal}}{E_0} & = & -0.5717 + 0.1416\log_{10} 
\frac{E_{\rm cal}}{\rm eV} \\ 
& &- 0.003328 \left(\log_{10}\frac{E_{\rm cal}}{\rm eV}\right)^2.
\end{eqnarray}
This formula reflects a quadratic fit to the results from a
purely protonic composition, which is suitable for the energy
range ($E \ge 10^{18}$~eV) of our present analysis \cite{2010PhRvL.104p1101A}. 
A heavier composition would require a slightly larger correction.

After fitting the shower profiles, 
we subject the processed data to additional quality cuts,
which primarily ensure that the position of $X_{\rm max}$ 
was within the FD's field of view (Table~\ref{tab:pcuts}).
The separate distributions of $E_0$ for cosmic-ray events 
seen by BRM and LR are shown in Figure~\ref{fig:ehist}. 
Additionally, we include the distribution of events present 
in the final data sets of both detectors. 
To avoid double-counting these ``stereo'' events 
(identified by trigger
times coincident within 200~$\mu$s), 
we assign them the geometric mean of the two FDs' independent
calculations of $E_0$. This also avoids biasing the 
results in favor of one detector or the other. 
  
\begin{table}[tb]
\centering
  \begin{tabular}{lcc}    
    Profile cut &&\\
    \hline
    Successful profile fit && \\
    First observed depth & $\ge$ & 150~g~cm$^{-2}$\\
    Last observed depth &$\le$ & 1200~g~cm$^{-2}$\\
    Extent of observed depth & $\ge$ & 150~g~cm$^{-2}$\\
    $X_{\rm max}$ bracketed && \\
  \end{tabular}
  \caption{\label{tab:pcuts}Cuts used to select good data after the 
  profile fitting stage of the Black Rock Mesa and Long Ridge monocular analysis. The slant depth
  of shower maximum $X_{\rm max}$ must be ``bracketed''
  by appearing within the observed portion of the track. 
  }

\end{table}

\section{\label{sec:aperture}Aperture calculation}
\begin{figure}[tb]
\includegraphics[width=3.4in]{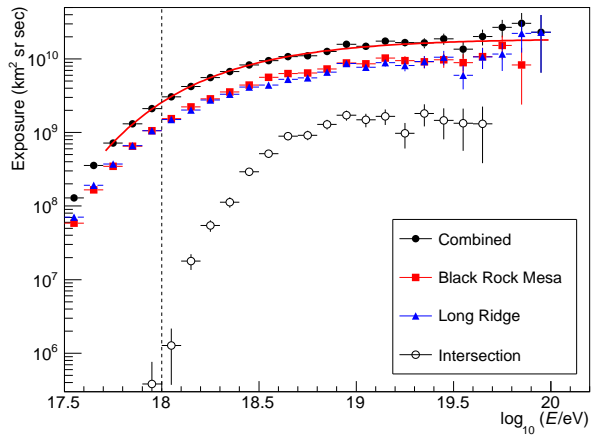} %exposure.eps
\caption{\label{fig:exposure}The calculated exposure for Black Rock Mesa
(red squares) and for Long Ridge (blue triangles). 
The two are summed bin-by-bin, and the intersection of the two exposures 
(open circles) is subtracted to calculate the combined exposure of the two
detectors, which is fit to Equation~\ref{eqn:exposure} (solid line). 
A dashed vertical line at reconstructed energy $E=10^{18}$~eV indicates the beginning of
our spectrum measurement.
}
\end{figure}

\begin{figure}[tb]
\includegraphics[width=3.4in]{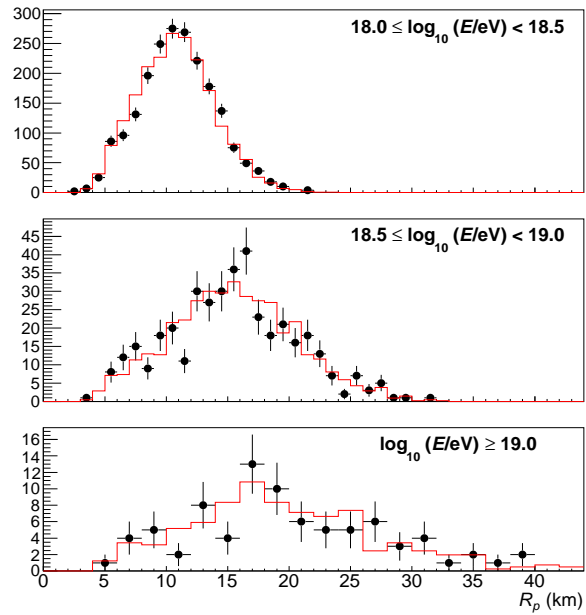} %brm_rp_dmc.eps
\caption{\label{fig:dmc_brrp}
Data/Monte Carlo comparison for Black Rock Mesa impact parameter $R_p$.
The comparison is divided into three ranges of reconstructed energy. 
Points represent data; the histogram represents Monte Carlo.
}
\end{figure}

\begin{figure}[tb]
\includegraphics[width=3.4in]{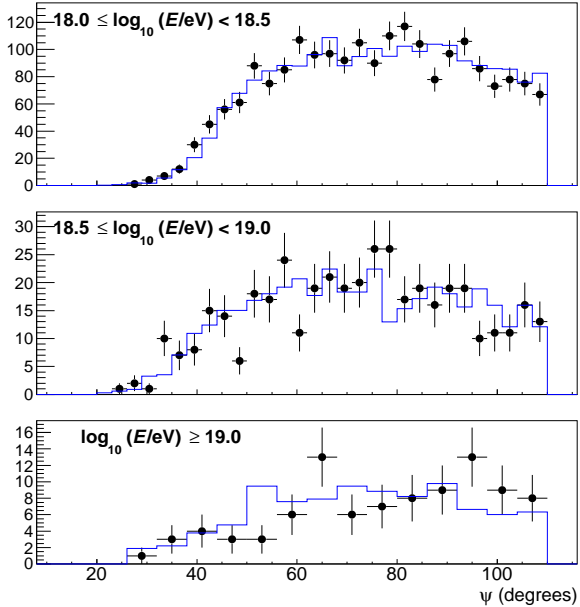} %lr_psi_dmc.eps
\caption{\label{fig:dmc_lrpsi}
Data/Monte Carlo comparison for Long Ridge shower inclination angle $\psi$.
Points represent data; the histogram represents Monte Carlo.
}
\end{figure}

\begin{figure}[tb]
\includegraphics[width=3.4in]{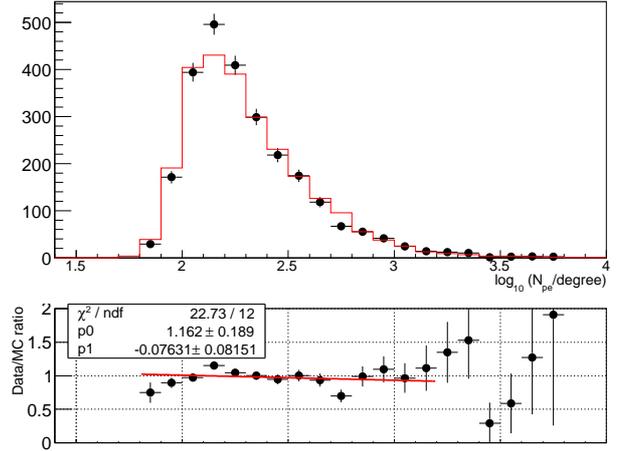} %brm_npedeg_dmc.eps
\caption{\label{fig:dmc_brnpd}
Data/Monte Carlo comparison for the number of detected photoelectrons per
degree of shower track at Black Rock Mesa. A linear fit to the ratio (only
considering bins containing a minimum of 10 data events) 
reveals no significant slope.
Points represent data; the histogram represents Monte Carlo.
}
\end{figure}

\begin{figure}[tb]
\includegraphics[width=3.4in]{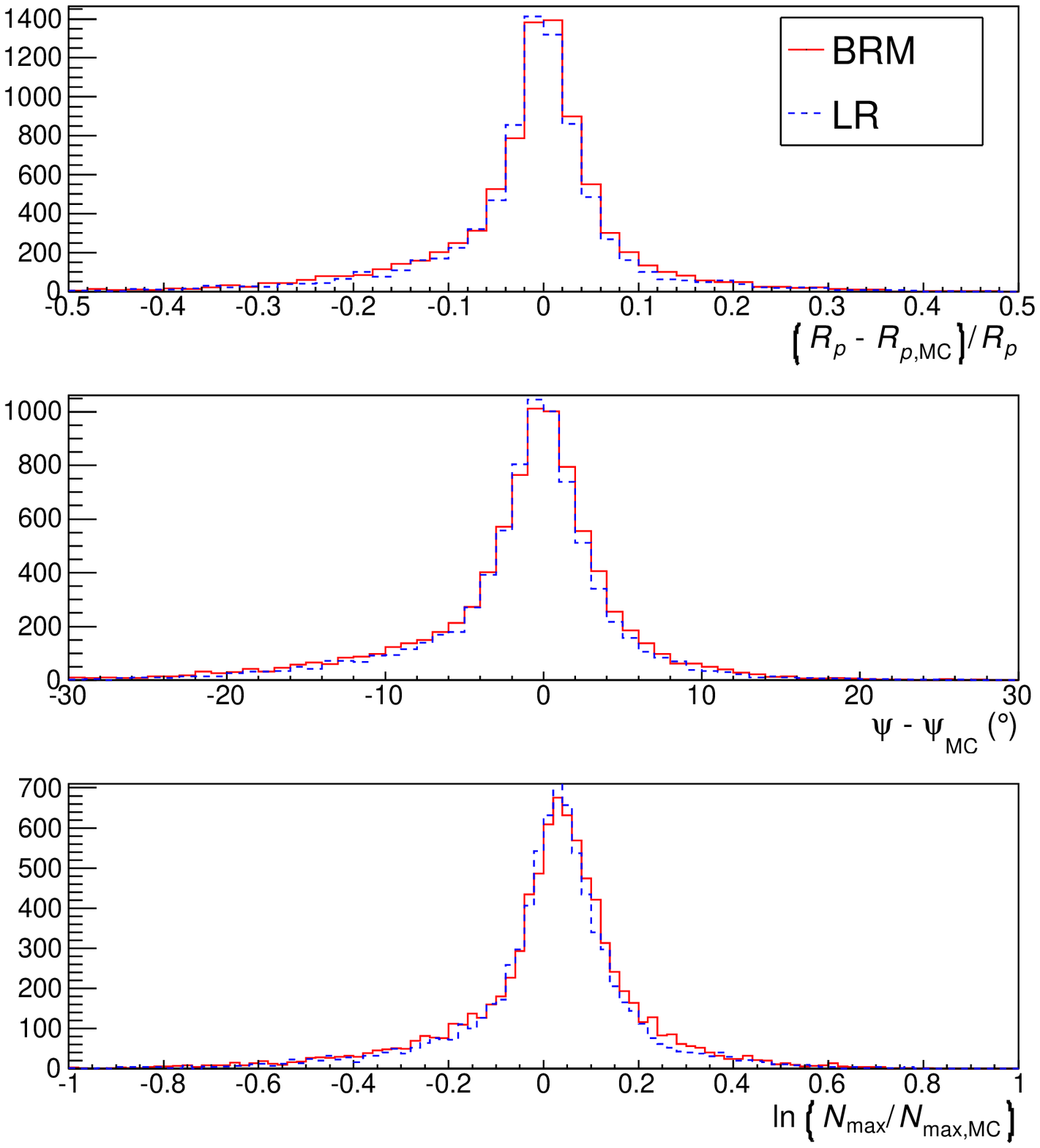} %res.eps
\caption{\label{fig:res}The detector resolution of Black Rock Mesa
(solid red lines) and Long Ridge (dashed blue lines) 
in two shower geometry observables (impact parameter $R_p$, top, and
inclination angle $\psi$, middle) and the Gaisser-Hillas profile parameter 
$N_{\rm max}$ (bottom). The respective FWHM values are 
7.7\%, 5.4$^\circ$, and 15.7\%.}
\end{figure}

The UHECR energy spectrum is related to each FD's data 
(Figure~\ref{fig:ehist}) by that detector's exposure, 
the subset of a multidimensional phase space in which 
cosmic rays of a given energy are detected and pass all quality cuts. 
In practice, this is the product of the
detector's live time and its energy-dependent aperture. 
The former is a straightforward calculation that subtracts 
dead time from the gross on-time of each detector as described in
Section~\ref{sec:fadc}. 
The latter is calculated by Monte Carlo simulation.

To minimize distortions caused by the finite ($\sim 10\%$)
energy resolution of monocular analysis, 
we simulate cosmic rays according to our best 
estimate of the true energy spectrum: 
the published HiRes spectrum \cite{2008PhRvL.100j1101A}
with a simulated proton fraction
based on the composition measurements by HiRes \cite{2010PhRvL.104p1101A} and HiRes/MIA \cite{2001ApJ...557..686A}. 
The randomly chosen impact points
for simulated showers are distributed uniformly in area 
over a region of Earth's surface 
(approximated as a sphere of radius 6370.98 ~km) 
within a radius of $1^\circ$ of arc (approximately 111~km) 
from the center of the TA experiment.
Each shower's local zenith angle is randomly selected 
between $0^\circ$ and $80^\circ$
according to an isotropic distribution. 
We choose the primary energy and particle type
(proton or iron) according to the aforementioned 
previous experimental results. 
A library of air showers previously generated in CORSIKA 
and binned by
energy and zenith angle for each particle species 
has many showers available per bin; 
the appropriate bin is picked, a shower is drawn at random, 
and its parameters are then scaled to match the desired
shower properties.

Because detector calibration and atmospheric conditions 
vary over the duration of the data set, 
we simulate cosmic rays for every data part 
according to its duration, so that the relative contributions of all
parts in simulation match those in data. 
We choose the mean rate of simulated showers so that the number of 
reconstruced Monte Carlo events passing all cuts 
is a few times larger than the corresponding number in the data. 
Given the parameter space in which we generate cosmic rays, 
the majority of our simulated showers do not trigger the
detector electronics. 
Those that are determined to be incapable of doing so based on
their energy and distance are simulated no further, 
but the rest advance to the ray-tracing simulation stage, 
which proceeds in the manner
described in the profile-reconstruction portion of 
Section~\ref{sec:analysis}, but without oversampling the detector acceptance. 
If a shower triggers the detector electronics, 
the simulated waveforms of all $3\sigma$ PMTs 
(including PMTs containing night-sky background only) 
from the triggered mirror(s) and any neighbors are
written to disk in a format identical to the preprocessed raw FD data.

The simulated-data files are then passed through the 
entire chain of data analysis described in Section~\ref{sec:analysis}. 
The aperture of the detector is calculated for an energy
interval (width 0.1 in $\log_{10} \left(E/{\rm eV}\right)$) 
by determining the ratio of reconstructed to simulated showers 
in that energy interval and multiplying by the phase-space
volume of simulated showers, 
approximately $1.18\times 10^5$~km$^2$~sr. 
Our finite energy resolution permits showers simulated in one 
energy interval to be reconstructed in another, 
but by simulating a realistic spectrum of cosmic rays, 
our aperture calculation accurately models this leakage.

The respective apertures of BRM and LR have some intersection, 
especially at primary energies above $10^{19}$~eV, 
where showers may trigger both detectors. 
When combining the FD sites' individual results, 
the combined exposure is the sum of the sites' exposures minus 
their intersection.
The latter is calculated using a modified version of the 
individual-FD Monte Carlo simulation that simulates 
both detectors simultaneously, 
and in which only reconstructed showers passing 
quality cuts in both detectors enter the numerator 
of the ratio against simulated showers 
(again with the geometric mean of their reconstructed energies). 
The detector live time for the combined
exposure is the total time during which cosmic rays were simulated, 
reduced by an amount corresponding to the product of the 
live-time fractions of the two FDs (dead time at each
detector is assumed to be independent for this calculation).

The separate and combined exposures,
as well as the size of the subtracted intersection, 
are shown in Figure~\ref{fig:exposure}.
To reduce bin-to-bin fluctuations of statistical origin, 
we fit the combined exposure $\xi\left(E\right)$ with the 3-parameter function
\begin{equation}
\label{eqn:exposure}
\log_{10} \xi\left(E\right) = 
 p_1 \left(1-\exp
\left[-\frac{\varepsilon - p_2}{p_3}\right]\right)
\end{equation}
where $\varepsilon\equiv\log_{10}\left(E/{\rm eV}\right)$. 
The fit range, 
$17.7 \le \varepsilon \le 20.0$,
begins approximately one third of a decade in energy below our
minimum energy.
The best-fit values of the parameters are 
given in Table~\ref{tab:expparm}.

\begin{table}[tb]
  \centering \begin{tabular}{lc}
    Parameter & Value
 \\ \hline    $p_1$: & $ 10.277 \pm 0.018 $\\ 
    $p_2$: & $ 16.734 \pm 0.022 $\\ 
    $p_3$: & $ 0.513 \pm 0.013 $\\ 
  $\chi^2$ / degrees of freedom: & $16.18 / 20$
  \end{tabular}
  \caption{\label{tab:expparm}The best-fit values for the exposure $\xi(E)$ in units of km$^2$~sr~sec (Figure \ref{fig:exposure}) using Equation \ref{eqn:exposure} in the range $17.7 \le \varepsilon \le 20.0$.}
\end{table}

Although our data set extends down to energies below $10^{17.5}$~eV
(see Figure~\ref{fig:ehist}), the aperture calculation has
the least systematic uncertainty above $10^{18}$~eV. 
At lower energies, the uncertainty in the aperture grows quickly due to variations in trigger efficiency with
atmospheric aerosol transparency. Additionally, the poorly known chemical 
composition of incoming cosmic rays introduces a systematic uncertainty
that grows rapidly with decreasing energy beginning near $10^{18}$~eV. 
Consequently,
we begin our spectrum measurement at $10^{18}$~eV, which is indicated
by a vertical line in Figure~\ref{fig:exposure}.

Because this exposure calculation has been performed 
entirely by simulation, 
it is necessary to verify that the simulation is an 
accurate representation of reality. 
To this end, we perform numerous data/Monte Carlo comparisons. 
An observable is chosen, 
such as the number of photoelectrons in a PMT or the 
shower impact parameter $R_p$, 
and the distributions of that observable
in data and simulation are compared to assess the similarity in shape.
A further quantitative comparison can be made by taking the
ratio of the distributions and fitting it to a straight line.

In Figures~\ref{fig:dmc_brrp} through~\ref{fig:dmc_brnpd}, 
we present a selection of
data/Monte Carlo comparisons from the present analysis. 
Figure~\ref{fig:dmc_brrp} shows the $R_p$ distributions broken into three
broad energy bands for the BRM FD. The good agreement between the data distribution
and the Monte Carlo distribution in all three bands
demonstrates the proper growth of the aperture with energy. 
Figure~\ref{fig:dmc_lrpsi} shows the $\psi$ distributions broken into the
same three energy bands, but for the LR FD. The good agreement between data and
Monte Carlo demonstrates
the proper angular acceptance of the aperture simulation. The complementary
comparisons for each are essentially identical. Finally, 
Figure~\ref{fig:dmc_brnpd} shows the brightness of accepted showers at BRM,
which demonstrates the accurate simulation of the 
detector photometric threshold.

The exposure simulation's accuracy has been demonstrated by data/Monte Carlo comparison.
It thus enables us to calculate our detectors' resolution in several observables. 
The resolutions in $R_p$, $\psi$, 
and $N_{\rm max}$
are presented in Figure~\ref{fig:res}.
In particular, the 
$N_{\rm max}$
resolution constrains the smallest physically meaningful energy interval
when constructing our data and exposure histograms.

\section{\label{sec:spectrum}Monocular FD energy spectrum}
\begin{figure}[tb]
\includegraphics[width=3.4in]{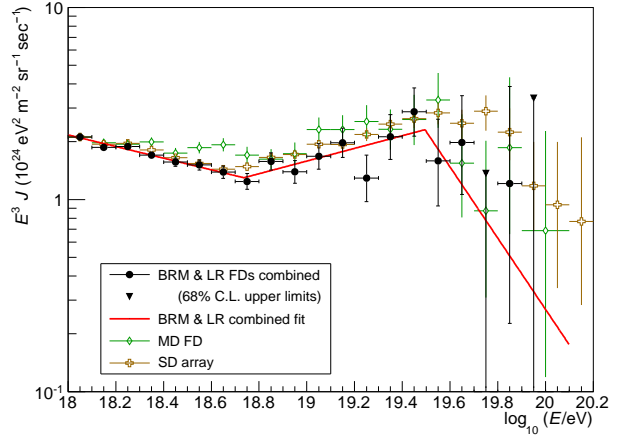} %spectrum.eps
\caption{\label{fig:fitspec}The spectrum from the data and
exposure of the combined Black Rock Mesa and Long Ridge FDs,
with a twice-broken power-law fit 
(parameters given in Table~\ref{tab:bplfit}). 
Included for comparison are two other spectrum measurements by
Telescope Array analyses: the Middle Drum FD (green diamonds) 
\cite{2012APh....39..109A}, 
and the surface detector array (brown crosses) \cite{2013ApJ...768L...1A}.}
\end{figure}

The energy spectrum of the UHECR flux $J\left(E\right)$ is the ratio of 
the number of data events to the exposure. 
In the $i^{\rm th}$ energy interval of width $\Delta E_i$ and where 
$E_i$ is the geometric mean of the interval limits, 
there are $n_i$ events, and
\begin{equation}
J\left(E_i\right) = \frac{n_i}{\xi\left(E_i\right)\Delta E_i}.
\end{equation}
For clarity's sake, we have multiplied the spectrum by $E^3$ 
in Figure~\ref{fig:fitspec}. For comparison, we also present
the published results from the 
MD FD and the SD ground array. Using a binned
maximum-likelihood fit, we determine parameters for the twice-broken
power law that best reproduces our data (Figure~\ref{fig:ehist}) given
our exposure (Figure~\ref{fig:exposure}). These parameters appear in
Table~\ref{tab:bplfit}.

\begin{table}[tb]
  \centering \begin{tabular}{lc}
    Parameter & Value
 \\ \hline    Flux at 1 EeV ($10^{-30} {\rm eV}^{-1} {\rm m}^{-2} {\rm sr}^{-1} {\rm sec}^{-1}$): & $ 2.17 \pm 0.05 $\\ 
    Power below ankle: & $ 3.30 \pm 0.03 $\\ 
    $\varepsilon$ at ankle: & $ 18.74 \pm 0.09 $\\ 
    Power above ankle: & $ 2.67 \pm 0.09 $\\ 
    $\varepsilon$ at cutoff: & $ 19.50 \pm 0.14 $\\ 
    Power above cutoff: & $ 4.9 \pm 0.9 $\\ 
  $\chi^2$ / degrees of freedom: & $11.92 / 15$
  \end{tabular}
  \caption{\label{tab:bplfit}The best-fit values for the combined spectrum (Figure \ref{fig:fitspec}) using a continuous twice-broken power-law function in the range $18.0 \le \varepsilon \le 20.1$.}
\end{table}

\section{\label{sec:combine}Combined FD monocular spectrum}
\begin{figure}[t]
\includegraphics[width=3.4in]{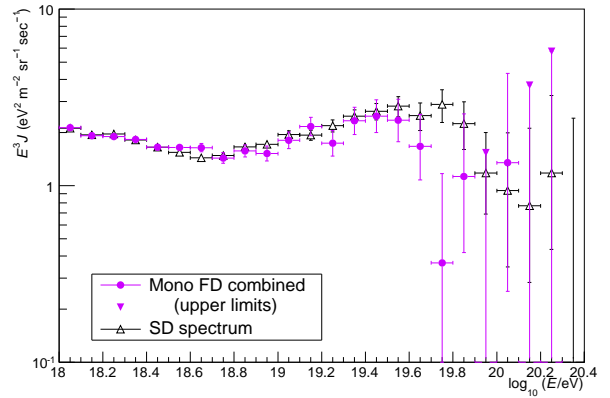} %combined.eps
\caption{\label{fig:tamono}The spectrum produced from the combined
data and exposure of the Black Rock Mesa, Long Ridge, and Middle Drum FDs.}
\end{figure}
To obtain a unified Telescope Array measurement of the UHECR spectrum in monocular mode, we combine our FADC-based FD result with the independent measurement
performed using the MD FD \cite{2012APh....39..109A}. 
To combine these measurements properly, we merge their observed
UHECR data sets (an event observed by more than one detector enter
the merged data set with the geometric mean of its measured energies)
and their respective exposures. Combining the exposures requires that
we account for their intersection, which we estimate via the
energy-dependent fraction of events in the MD data set seen
by at least one other detector. The resulting TA FD monocular spectrum
is shown in Figure~\ref{fig:tamono}, 
with the SD spectrum \cite{2013ApJ...768L...1A}.

\section{\label{sec:discussion}Discussion and conclusions}
The UHECR energy spectrum, as measured by TA's FADC-based FDs in
monocular mode using 3.5 years of data, 
is shown in Figure~\ref{fig:fitspec}. 
The shape of the spectrum plot is dominated by a power-law
dependence of flux on energy, punctuated by two abrupt changes 
in the spectral index. 
These breaks, a hardening of the spectrum near
$10^{18.74}$~eV and a softening near 
$10^{19.50}$~eV, are respectively recognizable 
as the ``ankle'' and a high-energy 
suppression likely associated with the Greisen-Zatsepin-Kuzmin (GZK) mechanism \cite{1966PhRvL..16..748G,1966JETPL...4...78Z}; both breaks have been
present in parallel and previous measurements of the 
UHECR spectrum 
\cite{2008PhRvL.100j1101A,2012APh....39..109A,2013ApJ...768L...1A,2008PhRvL.101f1101A}.
Our detection of the GZK suppression has a statistical significance of
$3.2\sigma$ relative to a spectrum that continues
unbroken from the ankle. We observe 
5 events above 
$10^{19.6}$~eV,
where an extension of the pre-GZK slope 
yields an expectation of approximately
16.88 events given the FDs' combined exposure.

The energy values at the spectral breaks 
are determined by our energy scale, 
which has a cumulative systematic uncertainty of 21\%.
Some of the larger contributors to this value are uncertainty in the
physics models used in calculating the calorimetric energy 
and fluorescence
yield as a function of primary energy and particle species (11\%) \cite{2008APh....29...77A}, 
as well as the attenuation of light by the 
atmospheric aerosol distribution (10\%; the observed RMS of the
VAOD distribution is 0.015, compared to the median value of 0.034 \cite{2011NIMPA.654..653T}). 
The absolute photometric calibration of the detectors 
contributes another 11\% systematic uncertainty 
to the energy measurement \cite{2012NIMPA.681...68K,2012.stratton.thesis}. 
Given the power-law nature of the spectrum, 
the 21\% systematic uncertainty on the energy 
results in an uncertainty on the
measured UHECR flux of 35\%. In Figure~\ref{fig:fitspec}, the
spectrum measurement from the FADC-based FDs is indeed
systematically lower than the spectra from the MD FD
and the SD array, but the difference 
is within our
systematic uncertainty\footnote{The MD FD systematic uncertainties (17\% energy scale, 30\% flux) are largely independent of those for the FADC FDs, having only the fluorescence-yield uncertainty in common.}.

\section{\label{sec:ackn}Acknowledgments}
The Telescope Array experiment is supported 
by the Japan Society for the Promotion of Science through
Grants-in-Aid for Scientific Research on Specially Promoted Research (21000002) 
``Extreme Phenomena in the Universe Explored by Highest Energy Cosmic Rays,'' 
and the Inter-University Research Program of the Institute for Cosmic Ray 
Research;
by the U.S. National Science Foundation awards PHY-0307098, 
PHY-0601915, PHY-0703893, PHY-0758342, PHY-0848320, PHY-1069280, 
and PHY-1069286 (Utah) and 
PHY-0649681 (Rutgers), and through TeraGrid resources provided by Purdue University and Indiana University \cite{Teragrid}; 
by the National Research Foundation of Korea 
(2006-0050031, 2007-0056005, 2007-0093860, 2010-0011378, 
2010-0028071, R32-10130);
by the Russian Academy of Sciences, RFBR
grants 10-02-01406a and 11-02-01528a (INR),
IISN project No. 4.4509.10, and 
Belgian Science Policy under IUAP VI/11 (ULB).
The foundations of Dr. Ezekiel R. and Edna Wattis Dumke,
Willard L. Eccles, and George S. and Dolores Dor\'e Eccles
all helped with generous donations. 
The State of Utah supported the project through its Economic Development
Board, and the University of Utah through the 
Office of the Vice President for Research. 
The experimental site became available through the cooperation of the 
Utah School and Institutional Trust Lands Administration (SITLA), 
the U.S.~Bureau of Land Management, and the U.S.~Air Force. 
We also wish to thank the people and the officials of Millard County,
Utah, for their steadfast and warm support. 
We gratefully acknowledge the contributions from the 
technical staffs of our home institutions as well as 
the University of Utah Center for High Performance Computing (CHPC).

\bibliography{fdmono-bibliography}

\end{document}